\documentclass{aa}  
\usepackage{natbib}
\usepackage[breaklinks=true]{} 
\usepackage{float}
\usepackage{graphicx}
\usepackage{amsmath}
\usepackage{txfonts}
\usepackage[final]{hyperref}
\hypersetup{
  colorlinks=true,   
  urlcolor=blue,     
  linkcolor=blue,     
  citecolor=blue
}
\begin{document}

   \title{ASASSN-13dn: A Luminous and Double-Peaked Type II Supernova}

   \subtitle{}

   \author{E. Hueichapán\inst{1,2}
          \and
          J. L. Prieto\inst{1,2}
          \and
          R. Cartier \inst{1}
          \and
          C. Contreras \inst{3}
          \and
          M. Bersten \inst{4,5,6}
          \and
          T. Moriya\inst{7,8,9}
          \and
          C. Kochanek \inst{10,11}
          \and
          B. J. Shappee \inst{12}
          }

   \institute{Instituto de Estudios Astrof\'{\i}sicos, Facultad de Ingenier\'{\i}a y Ciencias, Universidad Diego Portales, Avenida Ejercito Libertador 441, Santiago, Chile. \\
              \email{emilio.hueichapan@mail.udp.cl}
    \and
    {Millennium Institute of Astrophysics MAS, Nuncio Monse\~nor Sotero Sanz 100, Off. 104, Providencia, Santiago, Chile.}
    \and
    {Carnegie Observatories, Las Campanas Observatory, Casilla 601, La Serena, Chile}
    \and
    {{Instituto de Astrof\'isica de La Plata (IALP), CCT-CONICET-UNLP. Paseo del Bosque S/N, B1900FWA, La Plata, Argentina}}
    \and
    {Facultad de Ciencias Astronomicas y Geofisicas, Universidad Nacional de La Plata, Instituto de Astrofisica de La Plata (IALP), CONICET, Paseo del Bosque SN, B1900FWA, La Plata, Argentina}
    \and
    {Kavli Institute for the Physics and Mathematics of the Universe (WPI), The University of Tokyo, 5-1-5 Kashiwanoha, Kashiwa, Chiba, 277-8583, Japan}
    \and
    {National Astronomical Observatory of Japan, National Institutes of Natural Sciences, 2-21-1 Osawa, Mitaka, Tokyo 181-8588, Japan} 
    \and
    {Graduate Institute for Advanced Studies, SOKENDAI, 2-21-1 Osawa, Mitaka, Tokyo 181-8588, Japan} 
    \and
    {School of Physics and Astronomy, Monash University, Clayton, Victoria 3800, Australia}
    \and
    {Department of Astronomy, The Ohio State University, 140 W 18th Avenue, Columbus, OH 43210, USA}
    \and
    {Center for Cosmology and AstroParticle Physics, 191 W Woodruff Avenue, Columbus, OH 43210, USA}
    \and
    {Institute for Astronomy, University of Hawaii, 2680 Woodlawn Drive, Honolulu, HI 96822, USA}
    }

   \date{Received \today}

 
  \abstract
   {}
   {We present observations of ASASSN-13dn, one of the first supernovae discovered by ASAS-SN, and a new member of the rare group of Luminous Type II Supernovae (LSNe II). It was discovered near maximum light, reaching an absolute magnitude of M$_{v}$ $\sim$ -19 mag, placing this object between normal luminosity type II SNe and superluminous SNe
}
   {A detailed analysis of the photometric and spectroscopic data of ASASSN-13dn is performed. The spectra are characterized by broad lines, in particular the H$\alpha$ lines where we measure expansion velocities ranging between 14000 - 6000 km s$^{-1}$ over the first 100 days. H$\alpha$ dominates the nebular spectra, and we detect a narrow P-Cygni absorption within the broader emission line with an expansion velocity of 1100 km s$^{-1}$. Photometrically, its light curve shows a re-brightening of $\sim$ 0.6 mag in the $gri$ bands starting at 25$\pm$2 days after discovery, with a secondary peak at $\sim 73$d, followed by an abrupt and nearly linear decay of  0.09 mag d$^{-1}$ for the next 35 days. At later times, after a drop of 4 magnitudes from the second maximum, the light curves of ASASSN-13dn shows softer undulations from 125 to 175 days.}
   {We compare ASASSN-13dn with other LSNe II in the literature, finding no match to both light curve and spectroscopic properties. We discuss the main powering mechanism and suggest that interaction between the ejecta and a dense CSM produced by eruptions from an LBV-like progenitor could potentially explain the observations.}
   {}


   \maketitle
%



\section{Introduction}

Core-collapse supernovae (CCSNe) are the explosive end of a massive star (M$_{ZAMS} >$ 8 M$_{\sun}$) \citep{Filipenko97, Heger2003}. When the progenitor star retains its hydrogen envelope, the spectra are dominated by prominent Balmer lines from hydrogen. These hydrogen-rich supernovae (SNe) are classified as type II SNe. Based on their photometric properties, type II SNe are usually classified into two major sub-classes: type II-P and II-L SNe \citep{Barbon79}. If the light curve is characterized by a plateau of nearly constant luminosity after peak it is classified as SN II-P. On the other hand, light curves characterized by a linear decay after its peak luminosity are classified as SN II-L. Nonetheless, recent studies have shown that a continuous distribution better describes the wide variety of light curve shapes rather than two sub-classes \citep{anderson14}.

The spectroscopic properties of type II SNe define two additional sub-classes: Type IIb and type IIn. Type IIb SNe are stripped envelope SNe that retain a fraction of their hydrogen envelopes at the time of the explosion \citep{filippenko93}. They exhibit strong hydrogen and helium lines at early phases, while hydrogen lines dissapear and helium lines dominate at later phases. Type IIn SNe are characterized by persistent narrow emission lines with velocities < 1000 km/s, these narrow lines are produced by photoionized and unshocked circumstellar material \citep[CSM,][]{Schlegel90}. The narrow lines are frequently found on top of a broad spectral feature which is produced by the shocked ejecta and CSM. The SN emission is dominated by the interaction between the SN ejecta and the dense CSM.

The main powering mechanisms for regular type II SNe are the energy of the explosion deposited in the envelope during shock wave propagation and the decay of the radioactive elements, mainly $^{56}$Ni, generated during explosive nucleosynthesis \citep{branchwheeler_book}. After the photospheric phase, where the energy deposited in the envelope is released by recombination processes, the tail of the light curve is mainly powered by the energy deposition from the radioactive decay of $^{56}$Ni $\xrightarrow{ } ^{56}$Co $\xrightarrow{ } ^{56}$Fe. With full $\gamma$-ray trapping, the expected decline rate at the phase is 0.98 mag per 100 days \citep{Woosley89, Nadyozhin94}.

In the last two decades, a new class of transient phenomena has been identified. While regular CCSNe have an absolute magnitude range between $M_V \sim -$14 to $-$18.5 \citep{Li2011}, explosions showing peak absolute magnitude brighter than $M_V \leq -21$ mag were classified as Superluminous SNe \citep[SLSNe,][]{Qumby11}. However, this luminosity threshold is not definitive, \cite{DeCia18} shows that SLSNe are a distinct spectroscopic subclass sometimes overlapping in luminosity with the normal population of stripped-envelope SNe. 

Over the last few years, the luminosity gap between regular CCSNe and SLSNe has been gradually populated by the so-called Luminous Supernovae (LSNe). With peak absolute magnitudes $M_V <  \sim -19$, these objects appear to spectroscopically bridge the gap between regular CCSNe and SLSNe, suggesting a continuum in their spectroscopic properties \citep{Gomez22, Nicholl21}. While the discovery of type II supernovae within this luminosity range dates back several decades (e.g., SN 1979C,  \citealt{1979C}; SN 1998S, \citealt{1998S}; SN2003bg, \citealt{2003bg}), it was not until \cite{Pessi2023} that these kinds of explosions were formally recognized as a subgroup of type II SNe. Their study found that this subgroup evolves more slowly than typical type II SNe, exhibiting broader emission lines and ejecta velocities of around $v \sim10000$ km s$^{-1}$.

The light curves of CCSNe are generally characterized by a single peak. However, recent studies have increasingly identified double or multi-peaked light curves, particularly among stripped-envelope SNe \citep{Gutierrez21, Chen21, Cartier24}. These findings challenge traditional models and require the consideration of alternative powering mechanisms to account for the observations. Among the proposed central engine power sources are newly formed magnetars undergoing spin-down \citep{Kasen10, Woosley10} and fallback accretion onto a stellar-mass black-hole \citep{Dexter13}. These mechanisms have been invoked to explain extreme luminosities \citep{ogle14}, long-duration light curves \citep{iPTF14hls}, and distinct spectroscopic features observed in individual supernovae \citep[e.g.,][]{Nicholl15bn}. Additionally, the hypothesis of a double nickel distribution in stripped-envelope supernovae (SESNe) has provided a successful explanation for the observed light curves \citep{Bersten22}.

Due to the limited number of well-characterized multiple-peaked type II SNe, there is insufficient consensus regarding the mechanisms responsible for their powering. The observed properties, particularly the high luminosities, are difficult to reconcile with the typical powering mechanisms, suggesting that additional mechanisms are likely required \citep{Orellana18}.

An example of this debate is iPTF14hls \citep{iPTF14hls}, a type II SNe with long, luminous, and multiple-peaked light curves showing no clear signatures of ejecta-CSM interaction in the spectra. No narrow emission lines were observed in this event and an equal increase in luminosity at all wavelengths discarded the ejecta-CSM interaction scenario. Fallback into a stellar-mass black-hole was adopted as the leading powering mechanism to explain the bumpy light curve \citep{iPTF14hls}. \cite{Dessart18_14hls} argue in favor of a magnetar-powered model to explain the observed properties of this SN. 

A LSNe II with a similar discussion is the case of ASASSN-15nx \citep{15nx}. The linearity of its light curve might indicate that it is powered purely by radioactive decay with a significant gamma-ray leakage factor, but the $^{56}$Ni mass necessary to explain the peak luminosity is too large for CCSNe \citep{Muller17}. \cite{Chugai19} discuss the possibility of this SN being powered by a magnetar after modeling the light curve and the [\ion{O}{I}] 6300, 6364 \r{A} doublet in late-time spectra. 

Typically, a standard SN II is understood as the explosive endpoint of a red supergiant (RSG), and many of their observed characteristics align with this scenario. However, \citet{Morozova17} and \citet{Dessart22} argue that the continuum observed in light curve properties could be attributed to variations in CSM properties. In some cases, observed features such as the interaction of ejecta with fast-moving pre-explosion eruptions challenge the standard RSG scenario and suggest alternative progenitor types, including LBV-like stars. This is exemplified by events like SN~2009ip, a SN IIn, where pre-explosion eruptions and their interaction with the surrounding CSM indicate an LBV-like progenitor \citep{2009ip_spectra}.

In this paper, we report the unique luminous SN II ASASSN-13dn. With a peak absolute magnitude of $M_V \sim -19$ mag, the SN shows a secondary maximum 73 days after maximum light, and at 125 days after a drop of 4 magnitudes from the secondary maximum, the SN shows bumps. Spectroscopically, ASASSN-13dn displays broad hydrogen emission lines with ejecta velocities $\sim$ 14000 km s$^{-1}$. In section \ref{sec:data} we present the data collected, describe the reduction process, and briefly describe the host galaxy properties. In Section \ref{sec:analysis} we describe both photometry and spectroscopic data of ASASSN-13dn. Then, in Section \ref{sec:discussion} we discuss possible power sources and the ejecta-CSM interaction signatures found in the observations. Finally, in Section \ref{sec:conclusions} we present our conclusions. 

\begin{figure}[H]
        \centering
        \includegraphics[width=0.5\textwidth]{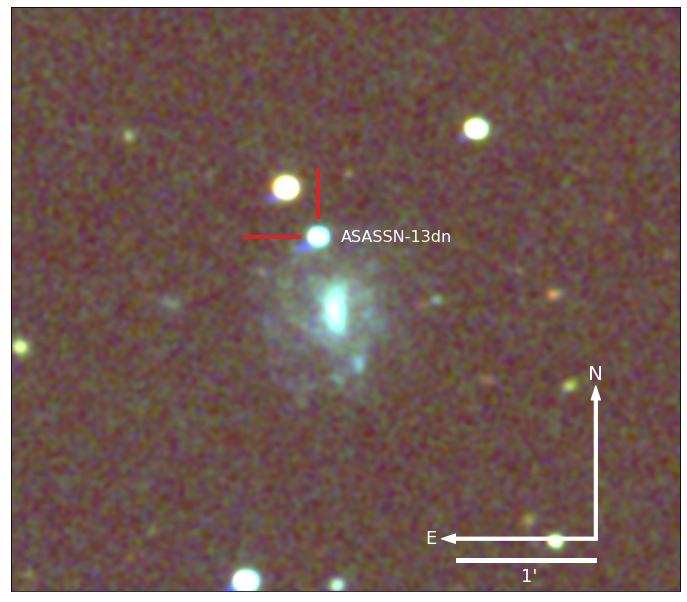}
        \caption{LCOGT $gri$ composite image of ASASSN-13dn. }
        \label{fig:finding_chart}
\end{figure}

\section{Data} \label{sec:data}

\subsection{Discovery and host galaxy}
\begin{figure*}[h!]
    \includegraphics[width=\textwidth]{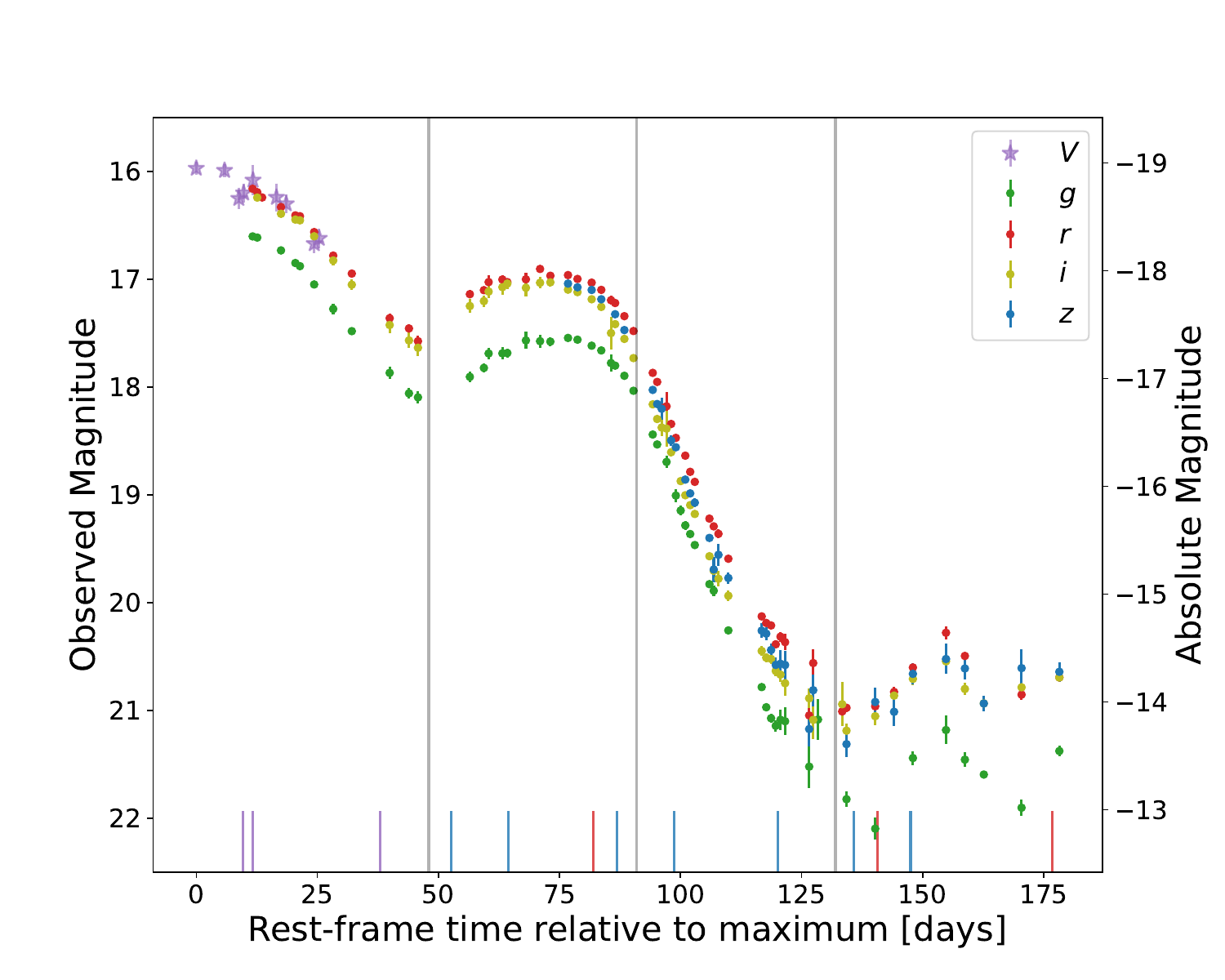}
    \caption{ASASSN-13dn ligth curve. A secondary peak is observed at $\sim$ 73 days after the first detection. In the tail of the light curve, clear sign of ejecta-CSM interaction are seen in the wiggles. The decline slope after the secondary peak is faster than the slope after the first peak. The lines at the bottom mark the epochs where spectroscopic data are available. The vertical lines divide the light curve in the phases described in section \ref{sec:LC}}
    \label{fig:LC}
\end{figure*}  

ASASSN-13dn was discovered by the All-Sky Automated Survey for Supernovae \citep[ASAS-SN;][]{Shappee14,Kochanek17} on December 5th 2013, 04:09 UT at a $V$-band magnitude of 15.70 by \cite{ATEL_Shapee} and on December 17th, 2013 at 21:03 UT the SN was classified as a type II SN \citep{AtelMartini}. It exploded in the outer regions of the star-forming galaxy SDSS J125258.03+322444.3 ($z=0.022636$) at $\alpha$=$12^{h}52^{m}58^{s}.20$, $\delta$ = $+32^{\circ}25'09".30$ (J2000.0). We adopt the Virgo Infall Only distance reported on NASA/IPAC Extragalactic Database of 96.0 Mpc ($\mu = 34.91 \pm 0.15$~mag) using a flat cosmology with $H_0$ = 73 km s$^{-1}$ Mpc $^{-1}$  and $\Omega_{m}$ = 0.27. At this distance, the supernova is $\sim 11.8$ kpc from the center of its host galaxy (see Fig \ref{fig:finding_chart}). We neglect any host-galaxy extinction due to the absence of any \ion{Na}{I} D absorption in all our spectra, indicating a very low or negligible contribution from the host. Therefore, we adopt a total reddening of $E(B-V)$ = 0.014 mag \citep{SF11}, assuming a R$_{V}$ = 3.1 \citep{Cadelli89} this results in A$_V$ = 0.043 mag.

\cite{Taggart21} include this galaxy in their photometric analysis of CCSNe hosts and estimate a total stellar mass of log$_{10}$(M$_{\star}$/M$_{\odot}$) = 9.85 $\pm$ 0.05 and a specific star-formation rate of log$_{10}$(sSFR/yr$^{-1}$) = $8.69 ^{+0.01}_{-0.46}$. We extracted a spectrum of the host galaxy with the Multi-Object Double Spectrograph (MODS) at the Large Binocular Telescope (LBT) on July 2nd, 2015 when the supernova had already faded. From the spectrum, we detect nebular emission lines of H$\alpha$, H$\beta$, [O~III]~5007~\AA, and [N~II]~6583~\AA. We used the scaling relation of \cite{Kennicutt98} to estimate a star-formation rate (SFR) of the host from the H$_\alpha$ luminosity of log$_{10}$(SFR) = 0.79 M$_{\odot}$ yr$^{-1}$. Using emission line ratios and the scaling relations presented in \cite{Marino13}, we estimate an oxygen abundance of 12+log(O/H) = 8.33 and 12+log(O/H) = 8.38 with the O3N2 and N2 methods, respectively. These values are consistent with 12+log(O/H) = 8.39 and 12+log(O/H) = 8.34 for the O3N2 and N2 indicators, the mean values of the oxygen abundances of the sample of core-collapse SNe from \cite{Thallis}.

\subsection{Photometry} \label{sec:phot}

\begin{figure}
    \centering
    \includegraphics[width=0.5\textwidth]{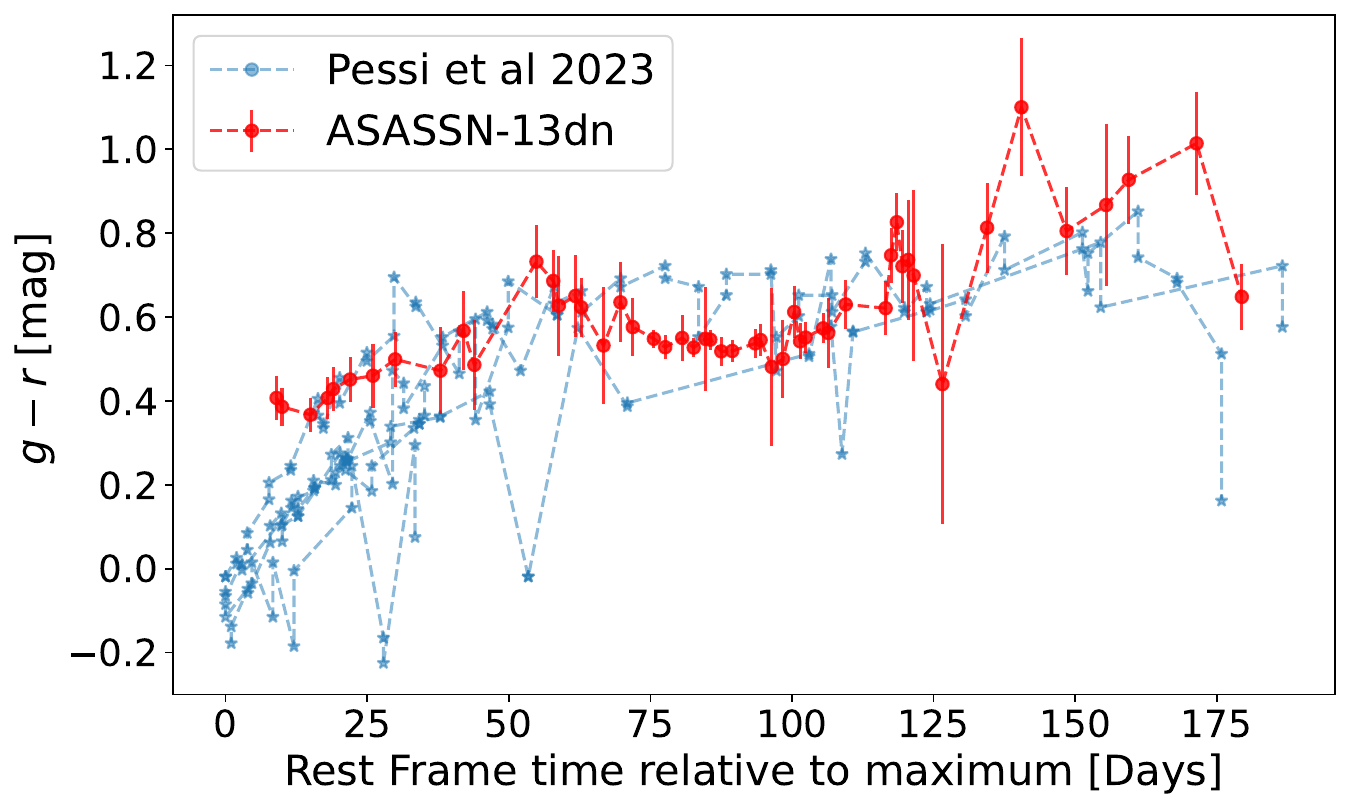}
    \caption{$g-r$ color evolution of ASASSN-13dn compared with LSNe II presented in \cite{Pessi2023}. Despite the variety of light curve shapes, the colors are consistent between objects.}
    \label{fig:colors}
\end{figure}  

\begin{figure*}
    \centering
    \includegraphics[width=\textwidth]{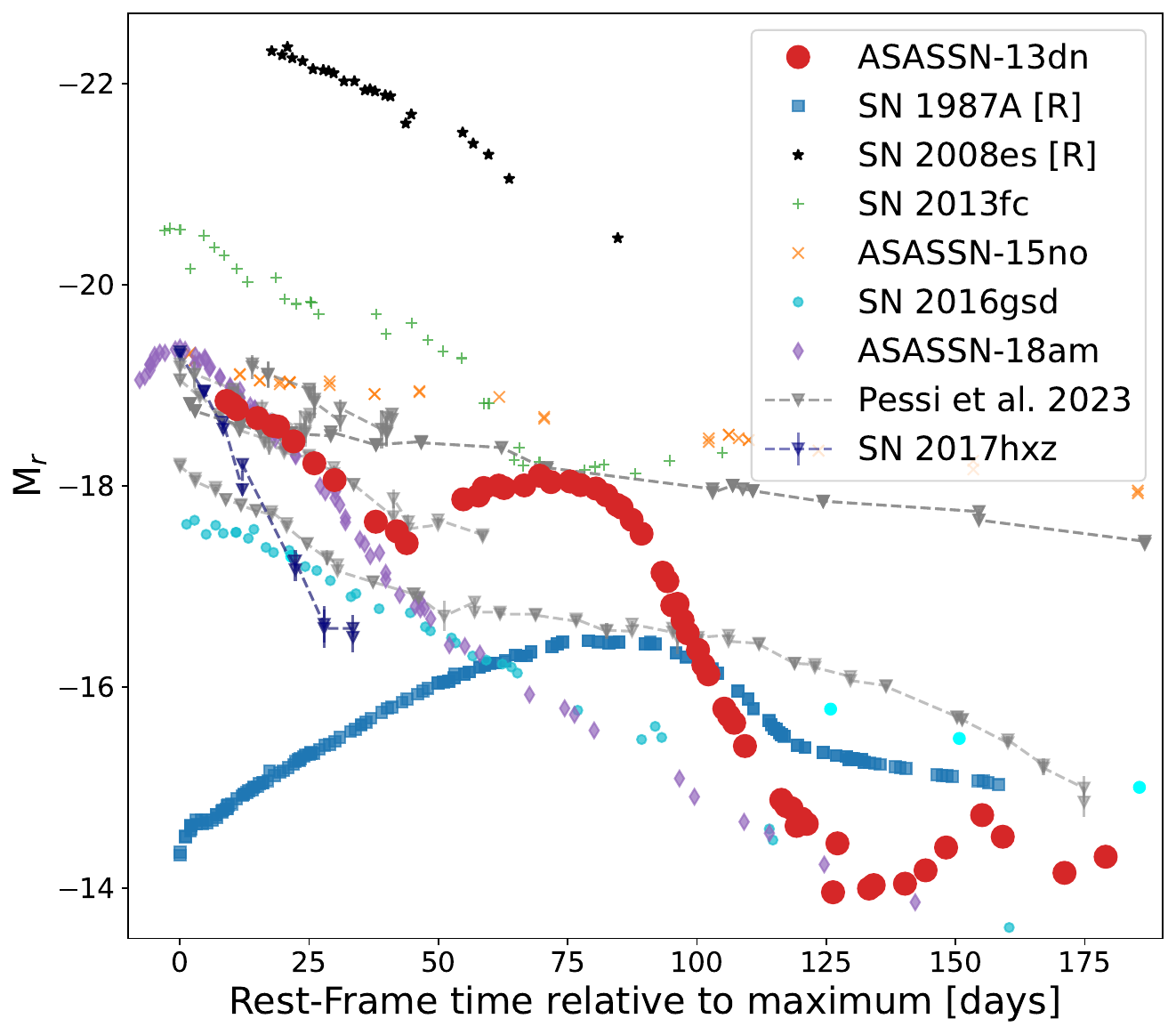}
    \caption{Comparison of the $r$-band absolute magnitude evolution with other LSNe II and SN~1987A. The phase is with respect to the maximum brightness. The bumps of ASASN-13dn are not seen in other objects. The adopted maximum light date in MJD, Distance (Mpc), and A$_V$ for each supernova are, respectively: 
    SN~1987A$^{1}$: 49849.8, 0.05, 0.206 \citep{87A}; 
    SN~2008es:      54599.3  1013.2 0.032  \citep{miller09}
    SN~2013fc:      56530.8, 83.2, 2.914 \citep{2013fc}; 
    ASASSN-15no:    57235.5, 153.5, 0.045 \citep{15no}; 
    SN~2016gsd:     57662.5, 311.6, 0.254 \citep{2016gsd}; 
    ASASSN-18am:    58142.6, 140.5, 0.027 \citep{15nx};
    SN~2017cfo:     57838.0, 178.2, 0.066;
    SN~2017hbj:     58031.0, 75.0, 0.095;
    SN~2017hxz:     58070.0, 330.6, 0.128;
    SN~2018aql:     58206.0, 321.4, 0.052;
    SN~2018eph:     58342.0, 121.8, 0.066
    \citep{Pessi2023}.\\
    \textit{1: Note that the phase reported for SN 1987A is relative to the explosion and not the maximum.}}. 
    \label{fig:r-comp}
\end{figure*} 




\begin{figure}
    \centering
    \includegraphics[width=0.5\textwidth]{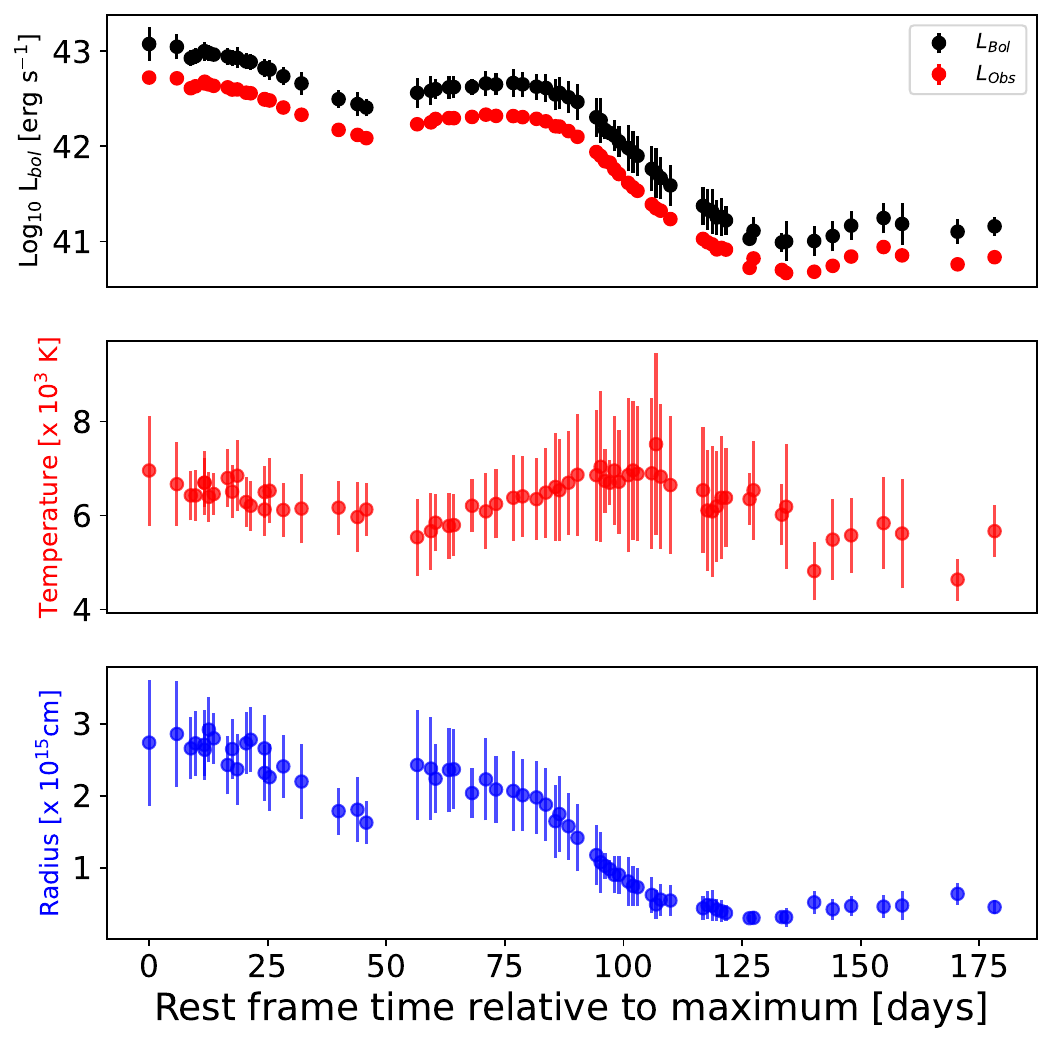}
    \caption{Bolometric light curve and best-fit parameters from the black-body fitting for ASASSN-13dn. The upper panel shows the bolometric and pseudo-bolometric light curve. The middle and bottom panels shows the effective temperature and blackbody radius derived from \texttt{superbol.py}, respectively. The increase at +40d might indicate an outward movement of the photosphere.}
    \label{fig:BolLC}
\end{figure}

The ASAS-SN $V$-band photometry using image subtraction is reported in Table \ref{Tab:V-Band}. The supernova was followed-up with the Liverpool Telescope \citep[LT;][]{LT2004} 2.0m and the Las Cumbres Observatory Global Telescope Network \citep[LCOGTN;][]{LCOGT13} 1.0m telescopes from discovery until $\sim 180$ days after the first detection. The $griz$ photometry was obtained performing PSF photometry with DoPhot \citep{Schechter93,AlonsoGarcia12}, using photometry of stars in the field from SDSS DR12 \citep{Alam15} to perform photometric calibrations. Image subtraction was performed on the $g$ and $r$ bands using a single archival SDSS DR12 image as a template. No major differences with the non-substracted photometry were seen, so no further attempts were made to improve the photometry. 

\begin{table}[H]
    \centering
    \begin{tabular}{cc}
    \hline
    \hline
    MJD & V \\
    (days) & (mag) \\
    \hline
    \hline
56632.10 & 15.97 $\pm$ 0.06 \\ 
56638.07 & 15.99 $\pm$ 0.06 \\ 
56641.08 & 16.25 $\pm$ 0.10 \\ 
56642.10 & 16.20 $\pm$ 0.09 \\ 
56644.08 & 16.08 $\pm$ 0.14 \\ 
56649.04 & 16.24 $\pm$ 0.13 \\ 
56651.09 & 16.30 $\pm$ 0.09 \\ 
56657.02 & 16.67 $\pm$ 0.09 \\ 
56658.08 & 16.62 $\pm$ 0.09 \\ 
    \hline
    \end{tabular}
    \caption{ASASSN-13dn $V$-band photometry.}
    \label{Tab:V-Band}
\end{table}

The SN exploded during a seasonal gap and no rise to peak was seen in the light curve. The last non-detection, on July 24.25 2013, 134d before the discovery \citep{AtelMartini}, so the explosion date is poorly constrained. In the first 10 days, the $V$-band light curve slowly faded and then transitions to a steeper decay rate. This suggests that the supernova was discovered shortly after maximum light. For simplicity, we assumed that the first detection is close enough to the epoch of maximum light, and all reported times are relative to the discovery.  

\begin{table*}
\centering
    \begin{tabular}{cccccc}
    \hline
    \hline
         MJD & g & r & i & z  \\
         (days) &(mag) &(mag)&(mag) &(mag) \\
         \hline
56644.01& 16.60 $\pm$ 0.03 &  16.16 $\pm$ 002 &  -- -- &  -- --\\
56644.98& 16.61 $\pm$ 0.02 &  16.19 $\pm$ 002 &  16.24 $\pm$ 002 &  -- -- \\
56646.01& -- --            &  16.24 $\pm$ 003 &  -- -- &  -- -- \\
56650.00& 16.73 $\pm$ 0.02 &  16.33 $\pm$ 001 &  16.39 $\pm$ 002 &  -- -- \\
56653.03& 16.84 $\pm$ 0.03 &  16.40 $\pm$ 002 &  16.44 $\pm$ 003 &  -- -- \\
... & ... & ...  & ... & ... \\

    \end{tabular}
    \caption{ASASSN-13dn $griz$ photometry from Liverpool Telescope (LT) and Las Cumbres Observatory Global Telescope Network (LCOGTN). The complete table can be found in the supplementary material.}
    \label{tab:photometry}
\end{table*}

\subsection{Spectroscopy} \label{sec:spec_red}

We obtained 13 optical spectra between +9d and +176d after maximum light using the Kitt Peak Ohio State Multi-Object Spectrograph \citep[KOSMOS;][]{Martini2014} on the KPNO Mayall 4m telescope, the Multi-Object Double Spectrograph \citep[MODS;][]{Pogge10} on the Large Binocular Telescope \citep[LBT;][]{Hill10}, and the Dual Imaging Spectrograph (DIS) on the Apache Point Observatory (APO) 3.5m telescope. The spectra were reduced using standard techniques in IRAF that included basic 2D image reductions (overscan and bias subtraction, flat-fielding), extraction of 1D spectra, wavelength calibration using HeNeAr lamps, and flux calibration using a standard star observed the night of the observation. We corrected the absolute flux calibration of the spectra by matching synthetic photometry from the spectra to the photometric light curve in the $r$-band.

\begin{figure*}[h!]
    \centering
    \includegraphics[width=\textwidth]{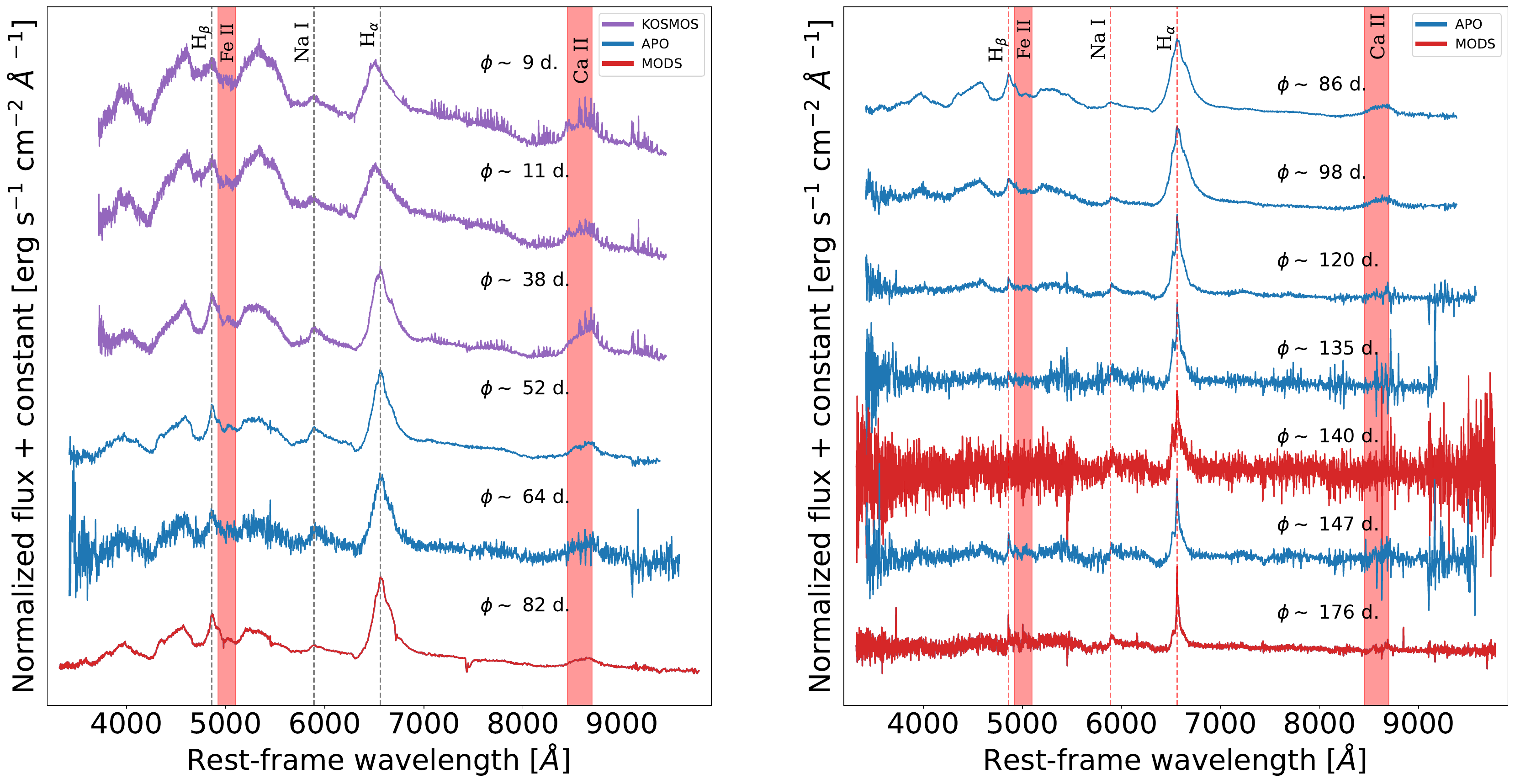}
    \caption{Spectroscopic evolution of ASASSN-13dn. The flux has been normalized to the maximum of H$\alpha$ for visibility purposes.}
    \label{fig:TimeSeries}
\end{figure*}

\section{Analysis} \label{sec:analysis}

\subsection{Light Curve} \label{sec:LC}

ASASSN-13dn was discovered at an absolute magnitude of $M_V = -18.94 \pm 0.21$ mag. We fitted a second-order polynomial to the $V$-band photometry, calculated the mean-fit polynomial derivative, and found that this value is constantly decreasing. As the first detection is the point where the derivative is close to zero, we assume that the maximum light is at MJD = 56631.2 $\pm$ 1.2. Using the $r$-band as a reference, we define four different sections of the light curve: 1) An initial decline from first detection to +48d; 2) A rise from +49d to a secondary peak at +73d. For the following days, it slowly decays to; 3) A second and stepper decline from +91d to +132d and; 4) A bumpy tail from +132d to the end of the observations. Fitting a quadratic polynomial to the $gri$ bands between +32d and +62d we find a minimum at +48.2$\pm$ 0.8 days (MJD = 56681.05) . The light curve has a nearly linear decay of $4.4$ mag/100d with a statistical error of $0.1$ mag after the first 10 days. We observe that the supernova reaches the secondary peak at different epochs for the individual $gri$-bands. Specifically, the later the peak occurs, the bluer the corresponding band (see Table \ref{tab:secondpeak}).

\begin{table}[]
    \centering
    \begin{tabular}{ccccc}
    \hline
    \hline
    Filter & MJD  &  Phase     &      $m$      &      $M$     \\
           & [days] &  [days] & [mag] & [mag]\\
    \hline
     g     & 56706.5  &  74.4 $\pm$ 2.8 &  17.55  $\pm$  0.03  & -17.36  $\pm$  0.03\\
     r     & 56705.1  &  73.0 $\pm$ 2.1 &  16.94  $\pm$  0.02  & -17.97  $\pm$  0.02\\
     i     & 56703.2  &  71.1 $\pm$ 2.6 &  17.04  $\pm$  0.04  & -17.87  $\pm$  0.04\\
        \hline
    \end{tabular}
    \caption{ASASSN-13dn secondary peak epochs, phases, and magnitude for individual $gri$ bands.}
    \label{tab:secondpeak}
\end{table}

The light curve brightens $\sim 0.6$ mag from +48d to +73d, and becomes bluer reaching its bluest color near the secondary peak. Next, the light curve decays 0.6 mag in $\sim 15$ days. This peak is followed by a faster decay of $9.3$ mag/100d with a statistical error of $0.1$ mag for the next 35 days. There is a second minimum at +132d and wiggles in the light curve can be observed. Figure \ref{fig:LC} shows the light curve of ASASSN-13dn, where the described phases can be clearly distinguished. 

The luminosity of this object place it into the group of hydrogen-rich luminous SNe. In Figure \ref{fig:r-comp} we present a comparison between ASASSN-13dn $r-$band light curve with SN~1987A and LSNe II from the literature. Immediately after maximum, most of the supernovae in Figure \ref{fig:r-comp} show a similar decay rate until $\sim 45$d, with the extreme exception of the fast-declining SN~2017hxz. After this phase, the variety of light curves makes them difficult to compare, nonetheless, linear decay at different phases is the common behavior and the secondary peak of ASASSN-13dn is unique among this group. After this, ASASSN-13dn's decay is faster than the other SNe presented here at similar phases. This is particularly clear when comparing the $r$-band light curve of ASASSN-13dn with ASASSN-15nx \citep{15nx}, an almost perfectly linearly declining SN II. 

Even though the light curves presented in Figure \ref{fig:r-comp} show a wide variety in their behavior, the $g-r$ colors (corrected for Galactic extinction) shown in Figure \ref{fig:colors} are remarkably consistent between LSNe II. A slight change towards the blue between +65d and +125d is detected, but it is still consistent with the blue colors of LSNe II. This change occurs at the same epochs where the secondary peak occurs. 

\subsection{Bolometric light curve}

We estimated the bolometric and pseudo-bolometric evolution using \texttt{superbol.py} \citep{nicholl18} to fit a modified blackbody to the photometry. The bolometric light curve integrates the model between 1500 and 25000 \r{A} while the pseudo-bolometric light curve only uses the wavelength range of the observed $griz$ photometry. Since it is the best sampled photometric band, we used the $r$-band detections as the reference light curve. The $z$-band was extrapolated assuming a constant $z-r$ color from peak to first $z$-detection at +65d. 
The resulting bolometric light curve, effective temperature, and blackbody radius are presented in Figure \ref{fig:BolLC}. We estimate a maximum bolometric luminosity of $1.3 \pm 0.6 \times 10^{43} $ erg s$^{-1}$ and a total radiated energy of $E_{tot} > 4.5 \times 10 ^{49}$ erg. 

An abrupt increase of 10$^{15}$cm in the radius at +55 days is observed and at the same time, a slow rise in the temperature of 2 $\times$ 10$^{3}$ K occurred. Simultaneously, the observed colors in Figure \ref{fig:colors} became bluer, consistent with the increasing temperature at the same epochs. This suggests an outward movement of the photosphere, potentially driven by an additional energy source boosting the light curves at this time. 

The wiggles at the end of the light curve are not consistent with a radioactive decay-powered explosion. Therefore we can only place an upper limit on the $^{56}$Ni synthesized in the explosion. Assuming a rise time between $15 - 25$ days and considering that the luminosity at +132d of 1.45 $\times$10$^{41}$ erg s$^{-1}$ is Nickel-powered with full trapping of gamma-rays, we used \cite{Nadyozhin94} to estimate an upper limit of $\sim 0.02$ M$_\odot$ of $^{56}$Ni, a factor of two lower than the mean nickel mass for a Type II SNe of M($^{56}$Ni) = 0.046 M$_\odot$ \citep{Muller17} and of M($^{56}$Ni) = 0.036 M$_\odot$ reported by \cite{Martinez22}.

\subsection{Spectra} \label{sec:spec}

Between +9d and +176d, 13 optical spectra of ASASSN-13dn were taken. The spectra are characterized by a broad H$\alpha$ emission line with velocities measured from the FWHM higher than 10000 km s$^{-1}$ until 86d post-maximum. Broad \ion{Fe}{II} at $\lambda$4924 $\r{A}$ and $\lambda$5018~$\r{A}$ lines are detected, but during the evolution of ASASSN-13dn those lines are blended with the H$\beta$ emission line and themselves, so no further analysis was done with the \ion{Fe}{II} features. \ion{Na}{I} at $\lambda$5889.9 $\r{A}$ and the Ca triplet at $\lambda$8498 $\r{A}$, $\lambda$8542 $\r{A}$ and $\lambda$8662 $\r{A}$ are also detected. \cite{Davis19} published a Near-Infrared (NIR) spectrum of ASASSN-13dn at +77d (MJD = 56711) that is mainly dominated by hydrogen, showing strong P$\gamma$ at $\lambda$1.094 $\mu$m, P$\beta$ at $\lambda$1.282 $\mu$m, and P$\alpha$ at $\lambda$1.875 $\mu$m. The spectroscopic evolution of ASASSN-13dn is presented in Figure \ref{fig:TimeSeries} and the spectroscopic log is in Table \ref{tab:spec_log}.

\begin{figure*}
    \includegraphics[width=\linewidth]{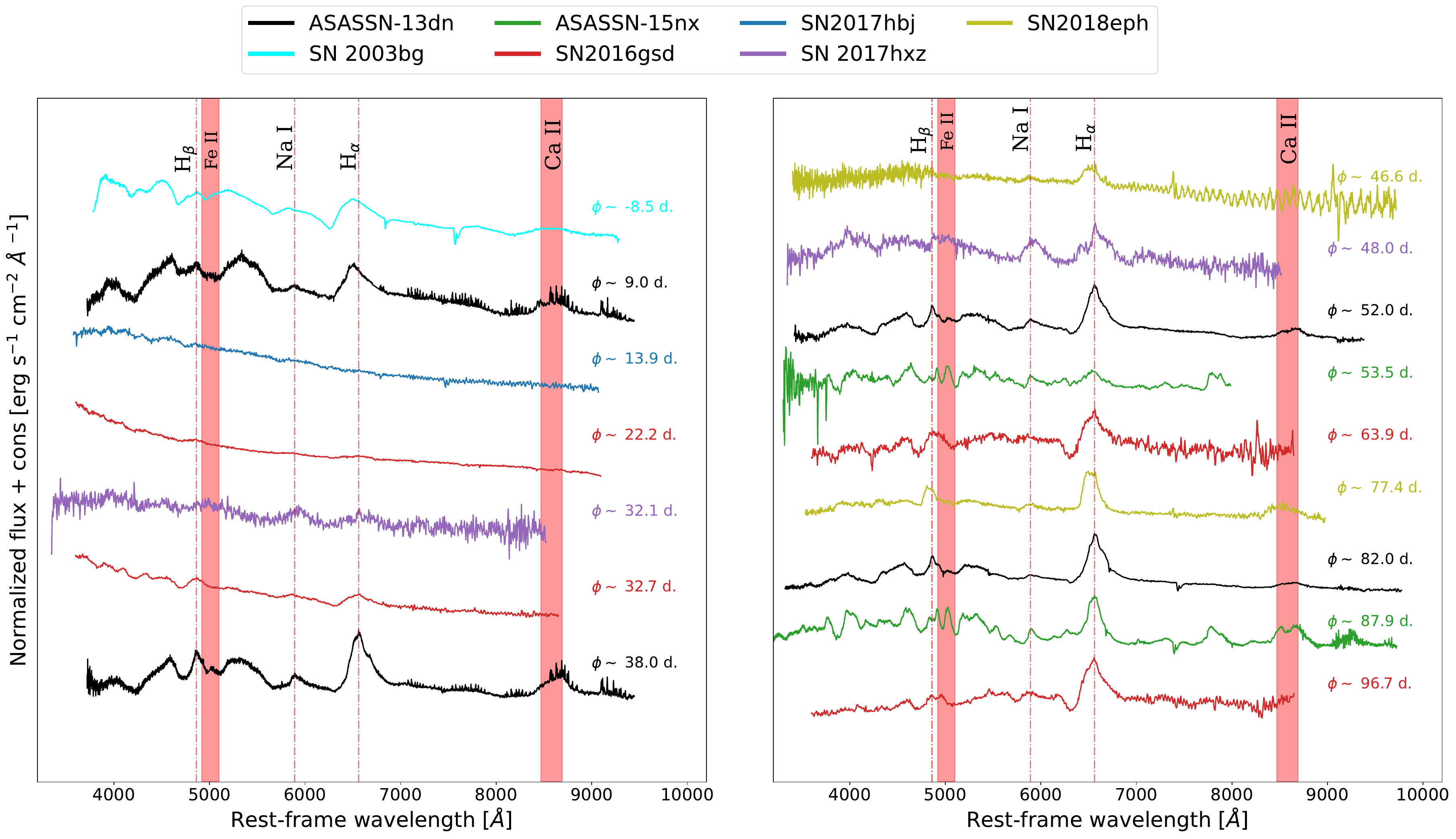}
    \caption{Comparison of ASASSN-13dn spectra with LSNe II studied in literature at different epochs. \textit{Left:} ASASSN-13dn peak spectra with SN~2003bg \citep{2003bg} ASASSN-15nx \citep{15nx}, SN~2016gsd \citep{2016gsd}. Spectra of SN~2017hxz, SN~2017hbj, SN~2018eph are from \cite{Pessi2023}. \textit{Right:} Comparison ASASSN-13dn secondary peak spectra with ASASSN-15nx \citep{15nx}, SN~2016gsd \citep{2016gsd}. Spectra of SN~2017hxz, SN~2017hbj, SN~2018eph are from \cite{Pessi2023}. Phases reported in this figure are relative to the maximum luminosity reported in the references.}
    \label{fig:Spec_comp}
\end{figure*}

 In Figure \ref{fig:Spec_comp} we compare the spectra from +9d to +38d with other early spectra of LSNe II. ASASSN-13dn shows a strong emission line of H$\alpha$ with a weak P-Cygni absorption, blended \ion{Fe}{II} lines, and weak \ion{Na}{I} emission profile. The broad hydrogen profile at this phase is only comparable with the type IIb SN~2003bg \cite{2003bg}. Before +30d, SN~2016gsd \citep{2016gsd}, SN~2017hbj \citep{Pessi2023}, SN~2017hxz \citep{Pessi2023}, and SN~2018eph \citep{Pessi2023} are mainly a blue featureless continuum. 

Between +50d and +100d, all the spectra in Figure \ref{fig:Spec_comp} are characterized by strong and broad Balmer emission lines. Boxy H$\alpha$ emission profiles are seen in the spectra of SN~2018eph and SN~2017hbj. A strong \ion{Ca}{II} triplet at $\lambda$8498 $\r{A}$, $\lambda$8542 $\r{A}$ and $\lambda$8662 $\r{A}$ is also present in SN~2017hbj and ASASSN-15nx \citep{15nx}. Both ASASSN-13dn and SN~2017hxz shown an asymmetry towards the red wing of the H$\alpha$ line, which is not seen in other supernovae. With the exception of SN~2016gsd \cite{2016gsd}, P-Cygni absorption at these phases is either weak or not detected. 

\begin{figure}[ht]
    \centering
    \includegraphics[width=0.49\textwidth]{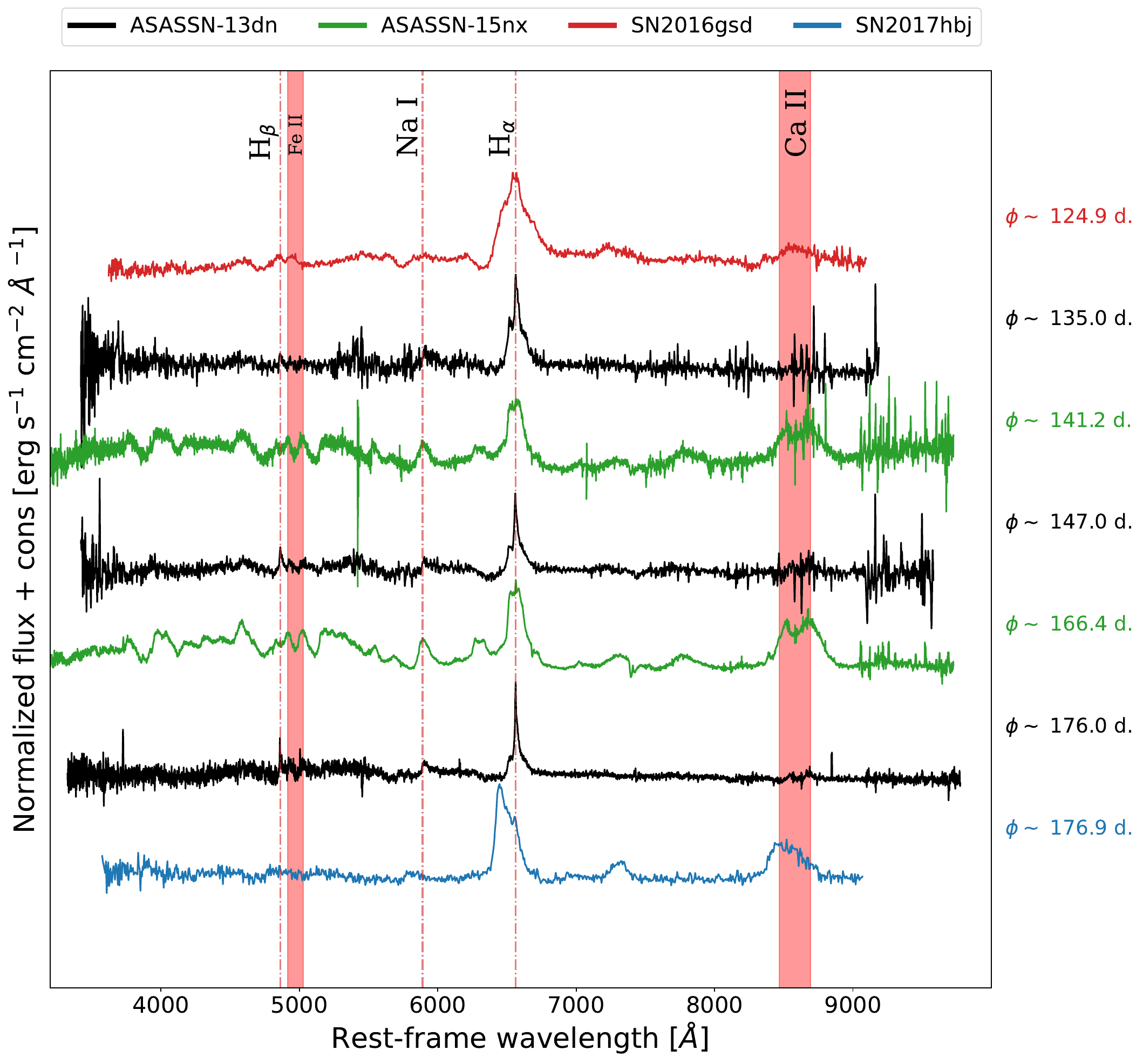}
    \caption{Comparison ASASSN-13dn late spectra with ASASSN-15nx \citep{15nx}, SN~2016gsd \citep{2016gsd} and SN~2017hbj \citep{Pessi2023}. Phases reported in this figure are relative to the maximum luminosity reported in the references.}
    \label{fig:Latespec}
\end{figure}

At later epochs, the spectra are dominated by Balmer emission lines with profiles that deviate from a single Gaussian component. Although the spectra of the other LSNe II shown in Figure \ref{fig:Latespec} are characterized by a broad and boxy H$\alpha$ emission line, the spectra of ASASSN-13dn are dominated by a narrow H$\alpha$ profile, with an absorption within the emission line at $\lambda$6540 $\r{A}$. \ion{He}{I} at $\lambda$7065 $\r{A}$, $\lambda$7281 $\r{A}$ and \ion{Ca}{II} $\lambda$8498 $\r{A}$, $\lambda$8542 $\r{A}$, and $\lambda$8662 $\r{A}$ lines are observed in nebular spectra of ASASSN-15nx and SN~2016gsd. Helium is not detected in ASASSN-13dn and a weak calcium line is detected in the last spectrum ($\phi +176$d).

\begin{table*}[!htbp]
    \centering
    \begin{tabular}{cccccc}
    \hline
    Date [UT] & MJD & Phase [d] (*) &  Instrument & Resolution & Range [$\r{A}$]\\
    \hline
      2013-12-16.97 & 56642.97  & 9    &    KOSMOS & 2800  & 3805 - 9660 \\ 
      2013-12-18.99 & 56644.99  & 11   &    KOSMOS & 2800  & 3805 - 9660 \\ 
      2014-01-14.03 & 56671.03  & 38   &    KOSMOS & 2800  & 3805 - 9660\\ 
      2014-01-29.78 & 56686.78  & 42   &    APO    & 800   & 3500 - 9800 \\ 
      2014-02-10.74 & 56698.74  & 64   &    APO    & 800   & 3500 - 9800 \\ 
      2014-02-28.87 & 56716.87  & 82   &    MODS   & 2000  & 3400 - 10000\\ 
      2014-03-05.75 & 56721.75  & 86   &    APO    & 800   & 3500 - 9800 \\ 
      2014-03-17.88 & 56733.88  & 98   &    APO    & 800   & 3500 - 9800 \\ 
      2014-04-08.87 & 56755.87  & 120  &    APO    & 800   & 3500 - 9800 \\ 
      2014-04-24.90 & 56771.90  & 135  &    APO    & 800   & 3500 - 9800 \\ 
      2014-04-29.69 & 56776.69  & 140  &    MODS   & 2000  & 3400 - 10000 \\ 
      2014-05-06.91 & 56783.91  & 147  &    APO    & 800   & 3500 - 9800 \\ 
      2014-06-05.68 & 56813.68  & 176  &    MODS   & 2000  & 3400 - 10000 \\ 
    \hline
    \hline
    \end{tabular}
    \caption{Log of spectroscopic observations. (*) : \textit{Phases are in rest frame and relative to the first detection. For simplicity were truncated to the unit.}}
    \label{tab:spec_log}
\end{table*}

\subsection{H$\alpha$ evolution}

\begin{figure}
    \centering
    \includegraphics[width=0.5\textwidth]{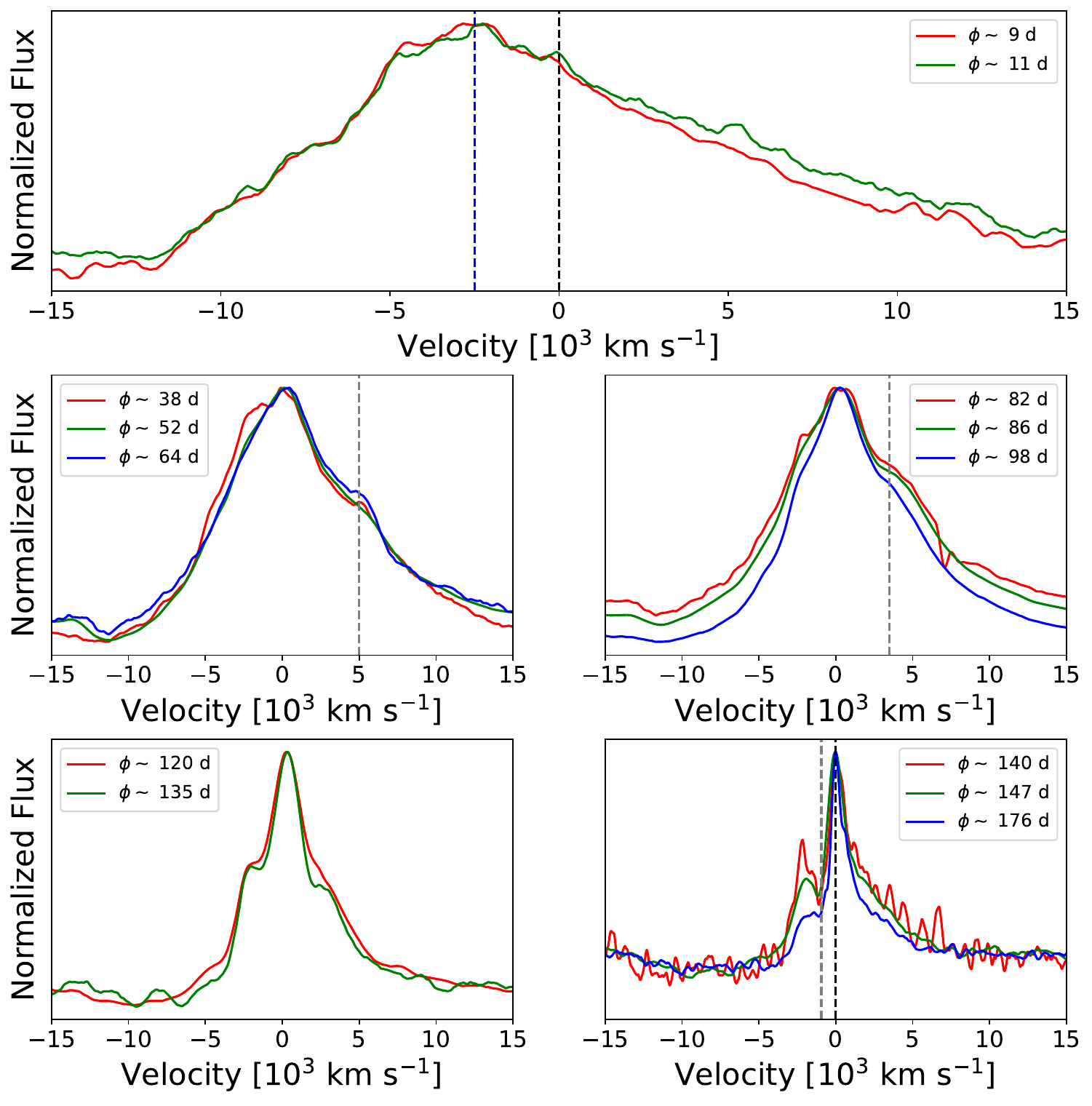}
    \caption{H$\alpha$ evolution of ASASSN-13dn. Top: Spectra close to the first detection, broad H$\alpha$ lines with velocities $v_{FWHM}$ close to 14000 km s$^{-1}$ and a blueshifted peak.  Middle panel: Persistent asymmetries towards the red part of the line from +38d to +98d at $\sim$5000 km s$^{-1}$. Bottom: Corresponding to the last epochs of the light curve H$\alpha$ shows asymmetries and a strong P-Cygni profile at +140d. This absorption has a velocity of $\sim$ 1100 km s$^{-1}$.} 
    \label{fig:halpha}
\end{figure}

The H$\alpha$ line profile and evolution shown in Figure \ref{fig:halpha} is complex. In the first two spectra, at +9d and +11d, the H$\alpha$ peak is blue-shifted by 2174 km s$^{-1}$. From +38d to +86d, a shoulder appears on the the red wing of the emission line. In the +98d spectra, the line shows bumps towards the red and the blue sections of the emission line and and finally, the blue bump seen at +98d evolves into a narrow P-Cygni profile detected between +135d and +147d. 

\begin{figure}
    \centering
    \includegraphics[width=0.5\textwidth]{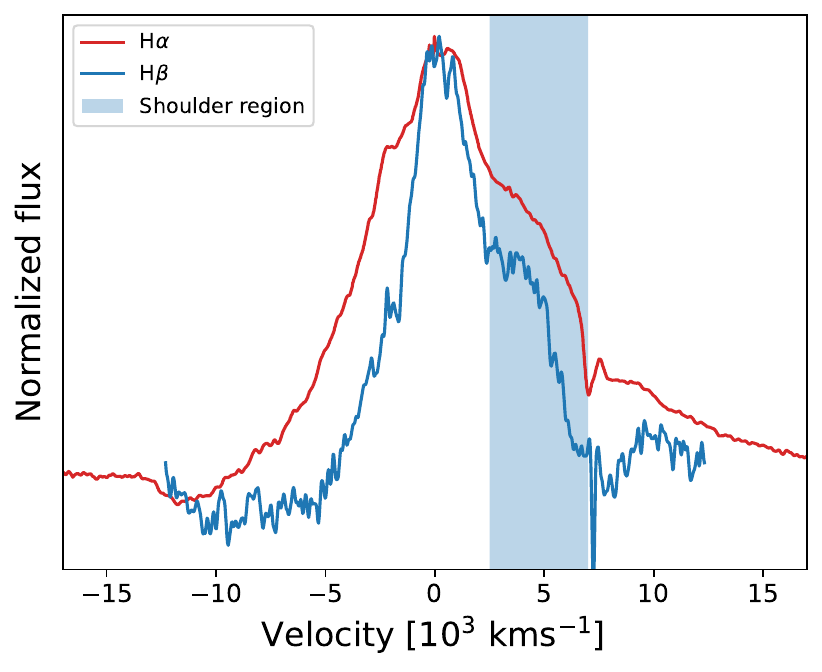}
    \caption{H$\alpha$ and H$\beta$ emission lines from the +82d spectrum. The blue region marks the shoulder detected in both H$\alpha$ and H$\beta$. The flux was normalized to the peak of the emission for visibility purposes.}
    \label{fig:HaHb}
\end{figure}

The blue-shifted peak of the emission observed in the +9d and +11d spectra has a velocity of 2174 km s$^{-1}$ and is consistent with that observed by \cite{Pessi2023} at later epochs for other LSNe II, nonetheless, it is not detected on the observations at +38d after peak. During the following observations, from +38d to +86d, a small but persistent shoulder appears on the red wing of H$\alpha$ at 6678 \r{A}. It is detected in every spectrum between +38d and +98d, being most prominent at +82d, coincident with the secondary peak in the light curve. A similar structure is present in the H$\beta$ line profile, particularly clear at +52d and +82d. Figure \ref{fig:HaHb} shows the emission profiles of H$\alpha$ and H$\beta$ at +82d where the shoulder is observed in both profiles with similar velocities, indicating that this structure is most likely related to asymmetries in the ejecta rather than the \ion{He}{II} line. 

\begin{figure}
    \centering
    \includegraphics[width=0.5\textwidth]{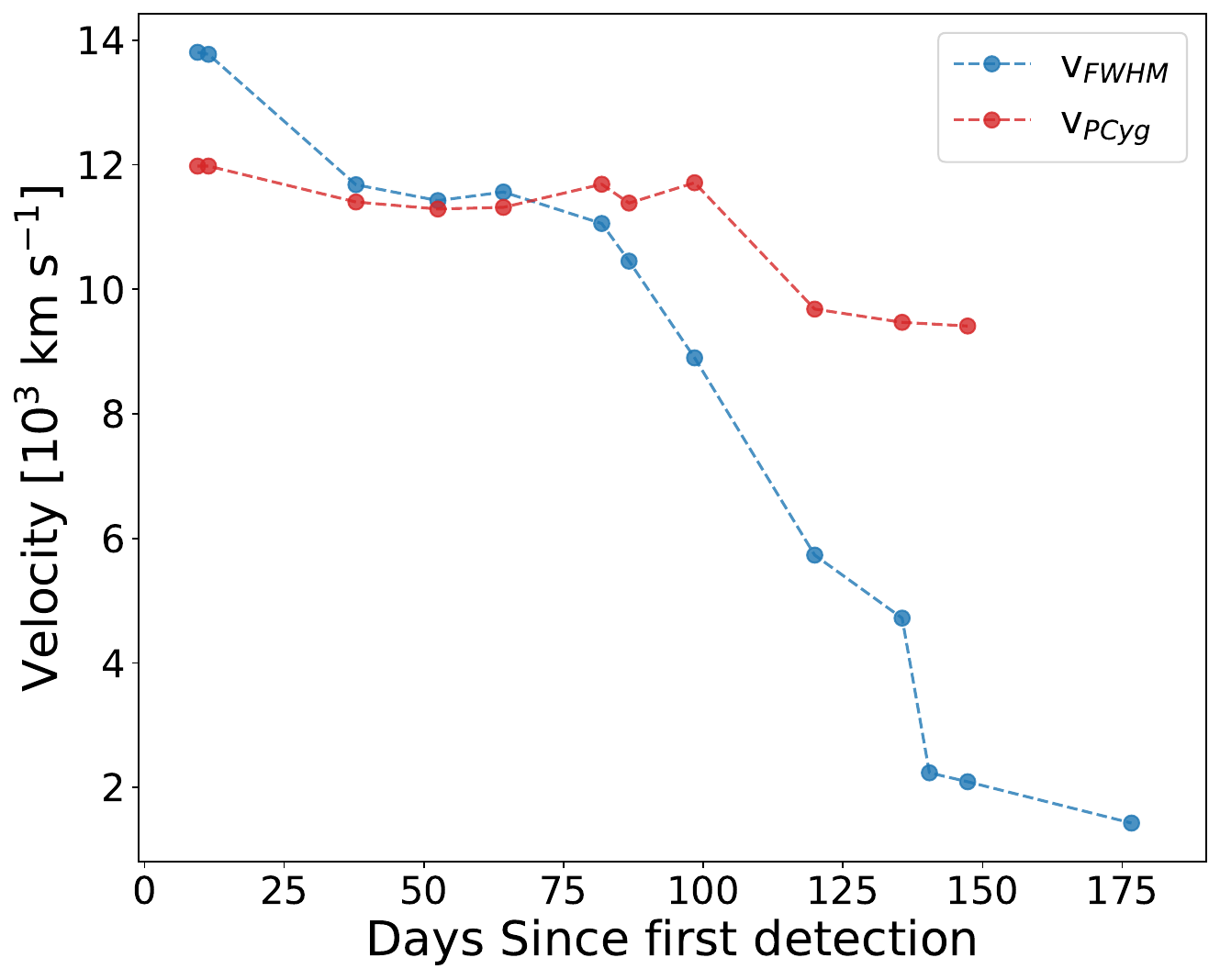}
    \caption{Velocity evolution of ASASSN-13dn. The $v_{FWHM}$ corresponds to the velocity measured from the FWHM of the H$\alpha$ emission and the $v_{Pcyg}$ is the velocity measured from the P-Cygni through.}
    \label{fig:velocity}
\end{figure}

We calculate the FWHM velocity ($v_{FWHM}$) and the velocity of the P-Cygni absorption trough ($v_{Pcyg}$) from the H$\alpha$ line in ASASSN-13dn. To fit and subtract the pseudo-continuum of the line, we used \texttt{PYTHON}'s package \texttt{specutils} \citep{specutils} in the range of the line profile for each spectrum. We also calculate the pseudo-equivalent width (pEW) of the line. This parameter quantifies the strength of the line compared to a pseudo continuum arbitrarily defined. In this case, we use the same \texttt{specutils} package to redefine a pseudo-continuum in the range of the H$\alpha$ emission. For both $v_{FWHM}$ and pEW we determine the uncertainties by randomly varying the wavelength range used to determine the pseudo-continuum by $\pm 5 \r{A}$ in a normal distribution. The evolution of the $v_{FWHM}$ and $v_{Pcyg}$ velocities are presented in Figure \ref{fig:velocity}.

\begin{figure}[h]
    \centering 
    \includegraphics[width=0.5\textwidth]{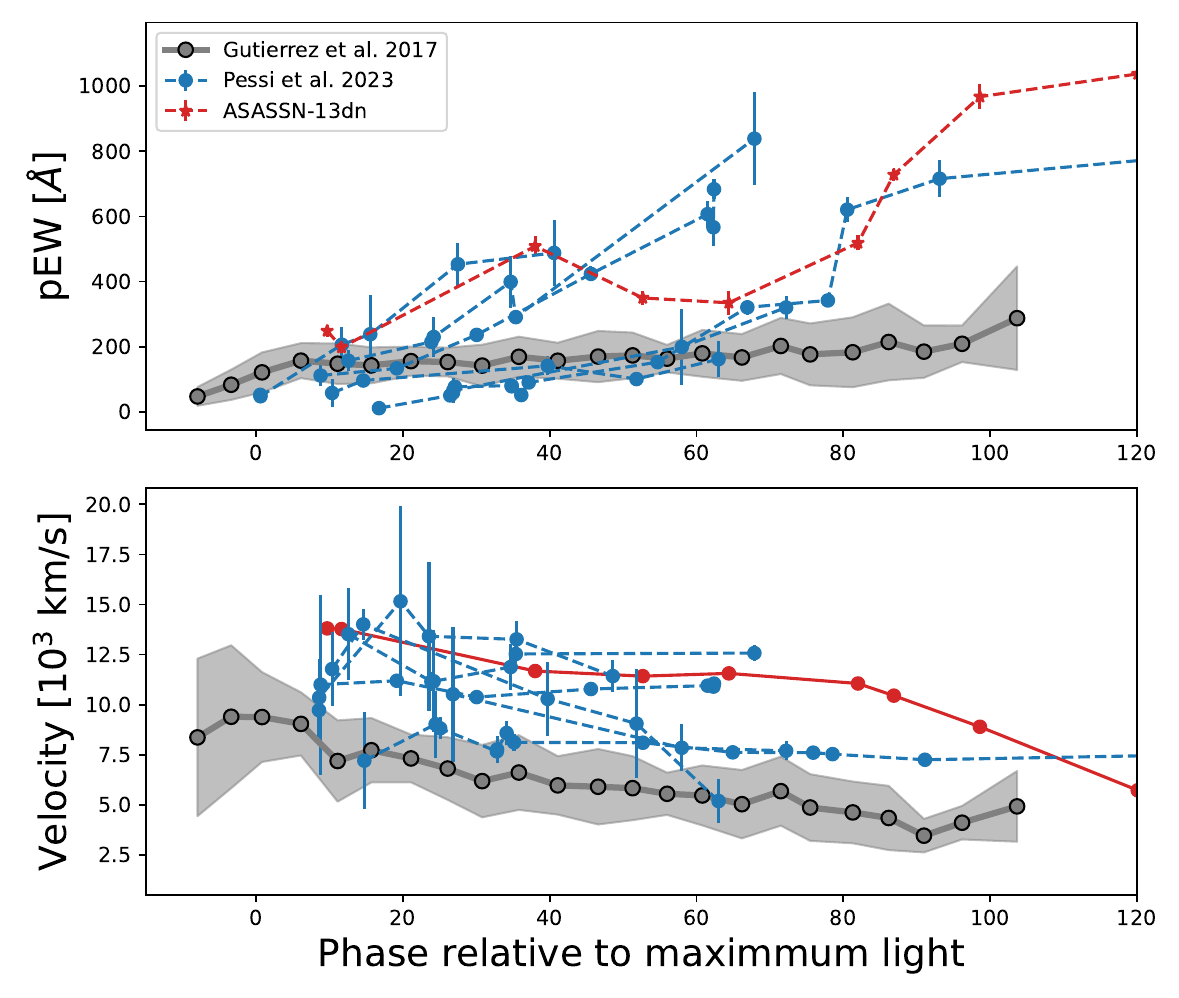}
    \caption{H$\alpha$ FWHM velocities ($v_{FWHM}$) and pEW of ASASSN-13dn compared with the sample of LSNe II from \cite{Pessi2023} and the SNe II sample from \cite{Gutierrez17}.} 
    \label{fig:pew}
\end{figure}

In figure \ref{fig:pew} we show the H$\alpha$ velocity measured from the FWHM ($v_{FWHM}$) and pEW evolution of ASASSN-13dn compared with the samples of normal type II SNe from \cite{Gutierrez17} and LSNe II from \cite{Pessi2023}. The velocity and pEW evolution of ASASSN-13dn is consistent with LSNe II. Between +9d and +98d, the P-Cygni velocities are always higher than 10000 km s$^{-1}$, but at +140d and +$176$d spectra, the absorption was not detected. The H$\alpha$ profile evolution is presented in Figure \ref{fig:halpha}. In the last three spectra, a P-Cygni profile inside the hydrogen line at $\lambda$6550 $\r{A}$ is detected, and a velocity of 1100 km s$^{-1}$ was measured.

\section{Discussion} \label{sec:discussion}

In this section, we explore different scenarios that could explain the luminosity and light curve features of ASASSN-13dn. Although the power source remains unclear, several signs of ejecta-CSM interaction are seen in the observations. The presence of a weak P-Cygni profile in the H$\alpha$ line has been interpreted as a consequence of interaction in type II SNe \citep{Gutierrez14}. 

A shoulder at $\sim$ 4100 km s$^{-1}$, detected in both the H$\alpha$ and H$\beta$ profiles near the second peak, is consistent with this interaction. Observed between $+38$d and $+86$d, it matches the rest wavelength of \ion{He}{I} emission, but it is most likely related to the ejecta-CSM interaction as it is also detected in the H$\beta$ line at a similar velocity (see Fig \ref{fig:HaHb}). A similar feature is described by \cite{Dessart22} in models and is associated with an increase in the interaction power of the model. 

In the later phases, the double-peaked H$\alpha$ profile and the bumpy light curve — particularly the wiggles observed after +128d — indicate a continued interaction between the ejecta and the surrounding material. The P-Cygni absorption profiles at +140d, +147d, and +176d, with a velocity of 1100 km s$^{-1}$, suggest interaction with a stellar wind or eruption from a massive progenitor star \citep{Smith14}. Similar profiles were observed by \cite{2009ip_spectra} in SN~2009ip, a type IIn SN, who proposed an LBV-like progenitor to explain the high velocity of the CSM shell. A comparable light curve signature, including similar wiggles, was observed in the type IIn SN iPTF13z, where the authors suggested that the interaction between the ejecta and denser CSM regions produced by LBV-like progenitor eruptions was responsible \citep{Nyholm17}.

In this scenario, three dense CSM shells at distances of $\sim 4.9 \times 10^{15}$ cm, $\sim 1.3 \times 10^{16}$ cm, and $\sim 1.6 \times 10^{16}$ cm, can produce the interaction signs at the secondary peak and subsequent wiggles observed in the light curve. These distances were determined by interpolating the H$\alpha$ FWHM velocity $v_{FWHM}$ (Fig~\ref{fig:velocity}) and integrating the interpolated velocities to calculate the distance traveled by the material at specific epochs. 
Assuming a rise time of 20 days and a constant velocity of 12000 km s$^{-1}$ up to day +9d, the secondary peak at +73d and the wiggles at +138d and +170d align with these shell locations. Assuming a velocity of the progenitor eruption/wind between 100 and 1000 km s$^{-1}$, the first shell is inferred to have been ejected between 25 to 2.5 years, prior to the explosion while the later shells, based on a CSM velocity of 1100 km s$^{-1}$ measured from the +140d H$\alpha$ spectrum \ref{fig:halpha} were produced 4.9 and 5.7 years before the explosion. 

Using these observations, the properties of the CSM can be constrained. The wind-density parameter is defined by $\Dot{M}$/$V_{CSM}$ = 2L/$V_{SN}^{3}$, where $V_{CSM}$ is the CSM velocity, $V_{SN}$ is the ejecta velocity, L the luminosity, and $\Dot{M}$ is the mass-loss rate. Using the bolometric luminosity of the first wiggle at +154d of 1.3 x 10$^{43}$ erg s$^{-1}$ and a mean velocity of 11000 km s$^{-1}$ from the P-Cygni absorption we calculate a wind-density parameter of $1.7\times 10 ^{16}$ g cm$^{-1}$. Additionally, with the CSM velocity measured at +140d we estimate a mass-loss rate $\Dot{M} \sim$ 0.08 M$_\odot$ yr$^{-1}$ in this scenario.

The light curve of ASASSN-13dn, shown in Figure \ref{fig:LC}, displays notable features, including a secondary peak and bumps at later times. In section \ref{sec:LC}, we measure the slopes of the light curve in different sections, finding a rapid decline immediately after the peak, followed by an even steeper decay after the secondary peak. The presence of CSM-ejecta interactions is known to enhance early-time luminosity, resulting in a rapidly declining light curve \citep{Morozova17}. The irregularities, particularly the wiggles observed post +128d, coincide with the appearance of the P-Cygni absorption at 1100 km s$^{-1}$ in the H$\alpha$ emission, further reinforcing the signature of ejecta-CSM interaction.

Energy injection by the spin down of a newly born magnetar \citep{Kasen10, Woosley10} has been proposed to explain the extreme luminosities of transients, particularly H-poor SLSNe \citep{Chatzopoulos13, Nicholl15bn, Hsu21}. This scenario has been proposed for a few hydrogen-rich SNe (iPTF14hls; \citep{Dessart18_14hls} , OGLE-2014-SN-073; \citep{Dessart_OGLE}, \citep{Orellana18}) as well as more normal stripped-envelope SNe \citep{rodriguez24}. In the case of ASASSN-13dn, a delayed power injection from a magnetar may boost the luminosity and explain the secondary peak.

The interaction between stars in a binary system has been proposed to explain pre-peak luminosity bumps, double-peaked light curves, and late-time wiggles \citep{DessartDP}. A recent example of a SN with late-time wiggles is SN~2022jli, a type Ic supernova showing periodic undulations in its light curve, that might be explained by the stripping of the outer envelope of a companion star \citep{Chen24, Cartier24}. This is in good agreement with the periodic displacement of the peak emission lines in the same timescale. Although the stripping of the outer layers of a companion could form a complex CSM structure which can explain the different structures in the emission lines of ASASSN-13dn, we cannot determine periodicity in the wiggles of the light curve. 

A mixture of several power sources is also a feasible scenario. \cite{Zhu24} proposed a model in which a magnetar formed in the SN can evaporate the outer layers of a companion star. The heating of the outer envelope can produce post-peak bumps and broaden the emission lines by accelerating the supernova ejecta. A similar scenario could explain the secondary peak, the wiggles, and the high ejecta velocities observed in ASASSN-13dn.

\section{Summary and conclusions} \label{sec:conclusions}

ASASSN-13dn is a unique example of a double-peaked luminous supernova in the LSNe II class. We present an analysis of this hydrogen-rich supernova, whose light curve exhibits a primary peak at $M_V \sim -19$ mag, placing it within the category of Luminous Supernovae. Approximately 45 days after reaching peak luminosity, the brightness of the supernova increases by 0.6 mag over a period of 30 days, reaching a secondary peak at +73 days. Then, the light curve shows a decline at a rate of $\sim$4.4 mag per 100 days. Finally, two distinct wiggles are observed in all bands from +125 days until the last detection.

Although ASASSN-13dn falls into the broad-line LSNe II group, some of the spectroscopic and photometric features differ from the sample shown by \cite{Pessi2023}. Most of the luminous supernovae are characterized by a blue continuum at early epochs while the peak spectra of ASSASN-13dn is dominated by strong hydrogen emission lines. This leads to a $g-r$ color at peak luminosity of ASASSN-13dn that is 0.4 mag redder than the LSNe-II \cite{Pessi2023}. This might indicate a problem with the phases calculated under the assumption that the first detection is the peak luminosity. 
Spectroscopically, ASASSN-13dn shows broad hydrogen emission lines from peak to +86d with measured FWHM velocities higher than 7000 km s$^{-1}$. The P-cygni through velocity remain fairly constant until +98d at $\sim$ 11000 km s$^{-1}$.
Finally, the P-Cygni through at 1100 km s$^{-1}$ in the last three spectrum seems to indicate that the ejecta is interacting with a fast-moving CSM probably produced by eruptions of an LVB-like progenitor.

\begin{acknowledgements}
E.H thanks Priscilla Pessi and Claudia Gutierrez for the data sent via private communication. E.H. was financially supported by Becas-ANID scholarship \#21222163 and by ANID, Millennium Science Initiative, AIM23-0001. J.L.P acknowledges support from ANID, Millennium Science Initiative, AIM23-0001. R.C. acknowledges support from Gemini ANID ASTRO21-0036

\end{acknowledgements}
%
%

\bibliographystyle{aa}
\bibliography{example} 

\end{document}